      [1999/12/01 v1.4c Il Nuovo Cimento]
\documentclass{cimento}

\usepackage{amsmath}
\usepackage{amsfonts} 
\usepackage{amssymb} 
\usepackage{amsthm} 
\usepackage{bbm} 
\usepackage{dsfont}
\usepackage{graphicx}
\usepackage{psfrag}
\usepackage{appendix}

\newcommand{\eq}[1]{Eq. (\ref{#1})}

\newcommand{\ii}{\mathrm{i}}
\newcommand{\ee}{\mathrm{e}}
\newcommand{\one}{{\rm 1\kern -.9mm l}}

\newcommand{\tr}{\mathrm{tr}\,}

\newcommand{\re}{\mbox{Re}\,}

\newcommand{\beq}{\begin{equation}}
\newcommand{\eeq}{\end{equation}}
\newcommand{\beqa}{\begin{eqnarray}}
\newcommand{\eeqa}{\end{eqnarray}}

\newcommand{\NN}{\mathcal{N}}

\title{Interplay between Background Fluxes and Instanton Configurations}
\author{Livia Ferro\from{ins:x}}
\instlist{\inst{ins:x}  LAPTH, Universit\'e de Savoie, CNRS\\
9, chemin de Bellevue, BP 110, 74941 Annecy-le-Vieux Cedex, France\\
livia.ferro@lapp.in2p3.fr}
\PACSes{PACS: 11.25.Uv, 11.25.Wx}

\begin{document}

\maketitle
\begin{flushright}
LAPTH-1371/09
\end{flushright}
\begin{abstract}
We focus on  D-brane models in presence of closed string background fluxes.  These fluxes modify the effective interactions on Dirichlet and Euclidean branes, in particular inducing fermionic masses. We show how they can create new non-perturbative superpotential terms in presence of gauge and exotic instantons in SQCD-like models.
\end{abstract}

\section{Introduction}
\label{intro}

In this paper we want to review how the presence of closed string background fluxes modifies the perturbative and non-perturbative sectors of the gauge theories realized by means of particular D-brane configurations.  
Background flux compactifications play many non-trivial roles in phenomenological models and represent an important ingredient of Type II string theories compactifications, in presence
of intersecting or magnetized D-branes preserving $\mathcal N=1$ supersymmetry. Indeed, they can create an effective potential for the moduli, break supersymmetry by generating soft supersymmetry breaking terms on D-branes and generate non-perturbative superpotentials in the low-energy theory. 
In particular, we will analyze
the possibility of adding internal (to preserve four-dimensional Lorentz invariance) fluxes both in the Neveu-Schwarz-Neveu-Schwarz (NSNS) and in the Ramond-Ramond (RR) sector of the bulk theory.

From the early studies, in the mid eighties, of heterotic string compactifications in presence of three-form H-flux \cite{Hull:1986iu}, flux compactifications have enormously developed.
The consequences of the presence of
internal NSNS or RR flux backgrounds onto the world-volume
theory of space-filling or instantonic branes
have been investigated relying on space-time supergravity methods \cite{Grana:2002tu}, and more recently on a string world-sheet approach \cite{Billo':2008sp}.
This world-sheet method is generic and allows to determine for instance the flux induced fermionic
terms of the D-brane effective actions through an explicit conformal
field theory calculation  of scattering amplitudes among string vertex operators. It can be applied also
to cases where the supergravity methods are less obvious, like
for instance to study how NSNS or RR fluxes couple to fields with twisted
boundary conditions, or how they modify the action which gives the measure of integration
on the moduli space of instantons.  Indeed, the interplay between fluxes and instantons is very deep. In presence of fluxes, non-perturbative superpotentials can be generated by instantons, giving rise to new low-energy effects \cite{Billo':2008pg}. Moreover, fluxes can contribute to get non-vanishing results in presence of exotic instantons by lifting fermionic zero-modes which would make vanish instanton-generated interactions.

This note is structured as follows. We first review three-form fluxes in $\NN = 1$ compactifications. We then present the basic steps to compute flux-induced couplings through world-sheet methods, focusing on the D3/D(-1) system. After a short review on instantons  in gauge and string theory, we give some explicit example of non-perturbative flux-induced superpotentials in the case of gauge and exotic instantons. In particular, we consider $\NN = 1$ SQCD, realized by fractional branes at a $\mathbb{C}^3/\big(\mathbb{Z}_2\times \mathbb{Z}_2\big)$ singularity.\\~\\
This paper is written for the INFN \emph{Sergio Fubini Prize} 2009, and is based on some results of the author's PhD thesis \cite{Ferro:2009gv}.

\section{Three-form Fluxes in $\NN = 1$ compactifications} 
\label{introflux}

As we already stressed in the Introduction, closed string background fluxes are an important ingredient  of D-brane models preserving $\mathcal N=1$ supersymmetry in Type II string theories compactifications.
These compactifications provide promising scenarios for phenomenological applications and realistic model building.  The effective actions of such brane-world models describe 
interactions of gauge degrees of freedom (associated to open strings) with gravitational 
fields (associated to closed strings), and have the generic structure of $\mathcal N=1$ supergravity 
in four dimensions coupled to vector and chiral multiplets. 
Four-dimensional ${\mathcal N}=1$ supergravity theories
are specified by the choice of a gauge group ${\mathcal G}$,
with the corresponding adjoint fields and gauge kinetic functions, by a K\"ahler potential $K$ 
and a superpotential $W$, which are, respectively, a real and a holomorphic function 
of some chiral superfields $\Phi^i$. 
The supergravity vacuum is parametrized by the expectation values of these chiral
multiplets that minimize the scalar potential
\begin{equation}
V\,=\,\ee^{K}\left( D_i \bar W D^i W - 3 \,|W|^2\right) +D^a D_a
\end{equation}
where $D_i W\equiv\partial_{\Phi^i} W + \big(\partial_{\Phi^i} K \big)\,W$ is the K\"ahler
covariant derivative of the superpotential and the $D^a$ ($a=1,\ldots,{\rm dim}(\mathcal G)$)
are the D-terms. Supersymmetric vacua, in particular, correspond to those
solutions of the equations 
$\partial_{\Phi^i} V=0$  satisfying the D- and F-flatness conditions $D^a=D_i W=0$.
The chiral superfields $\Phi^i$ of the theory
 comprise the fields $U^r$ and $T^m$ that
parameterize the deformations of the complex and K\"ahler structures of the three-fold,
the axion-dilaton field
\begin{equation}
\tau=C_0+\ii \,\ee^{-\varphi}~,
\end{equation} 
where $C_0$ is the RR scalar and $\varphi$ the dilaton, and also some multiplets
$\Phi_{\mathrm{open}}$ coming from the open strings attached to the D-branes.
The resulting low energy ${\mathcal N}=1$ supergravity model has a highly degenerate vacuum.
One way to lift (at least partially) this degeneracy is provided by the addition
of internal three-form fluxes of the bulk theory
via the generation of a superpotential
\begin{equation}
W_{\mathrm{flux}}=\int G_3\wedge \Omega~,
\label{wflux3}
\end{equation}
where $\Omega$ is the holomorphic $(3,0)$-form of the Calabi-Yau three-fold and 
\beq
G_3={F}-\tau H
\eeq
is the complex three-form flux given in terms of the RR and NSNS fluxes  
${F}$ and $H$. The flux superpotential (\ref{wflux3}) depends explicitly on $\tau$ through $G_3$ 
and implicitly on the complex structure parameters $U^r$ which specify $\Omega$, while it does not depend on K\"ahler structure moduli $T^m$.
Using standard
supergravity methods, F-terms for the various compactification moduli can be obtained from (\ref{wflux3}). 
The F-terms can also be interpreted as
the $\theta^2$ ``auxiliary'' components of the kinetic functions for the gauge theory
defined on the space-filling branes,
and thus are soft supersymmetry breaking terms for the brane-world
effective action. These soft terms have been computed in various scenarios of flux compactifications and their effects, such as flux-induced masses for the gauginos and the gravitino,
have been analyzed in various scenarios of flux compactifications
relying on the structure of the bulk supergravity Lagrangian and on
$\kappa$-symmetry considerations (see for instance the reviews \cite{Grana:2005jc}); here we will briefly present their derivation through a direct world-sheet analysis.

\section{Flux interactions on D-branes from string diagrams}
\label{sec:CFT}

\begin{figure}[ht]
\begin{center}
\begin{picture}(0,0)%
\includegraphics{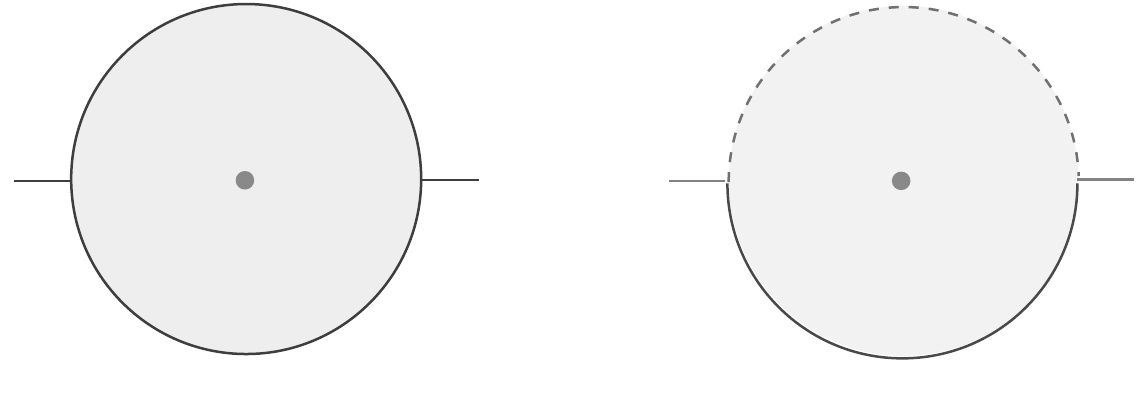}%
\end{picture}%
\setlength{\unitlength}{1579sp}%
\begingroup\makeatletter\ifx\SetFigFontNFSS\undefined%
\gdef\SetFigFontNFSS#1#2#3#4#5{%
  \reset@font\fontsize{#1}{#2pt}%
  \fontfamily{#3}\fontseries{#4}\fontshape{#5}%
  \selectfont}%
\fi\endgroup%
\begin{picture}(13638,4765)(271,-3920)
\put(286,-1825){\makebox(0,0)[lb]{\smash{{\SetFigFontNFSS{10}{12.0}{\familydefault}{\mddefault}{\updefault}$V_\Theta$}}}}
\put(5596,-1810){\makebox(0,0)[lb]{\smash{{\SetFigFontNFSS{10}{12.0}{\familydefault}{\mddefault}{\updefault}$V_{\Theta^\prime}$}}}}
\put(8146,-1823){\makebox(0,0)[lb]{\smash{{\SetFigFontNFSS{10}{12}{\familydefault}{\mddefault}{\updefault}$V_\Theta$}}}}
\put(13456,-1808){\makebox(0,0)[lb]{\smash{{\SetFigFontNFSS{10}{12.0}{\familydefault}{\mddefault}{\updefault}$V_{\Theta^\prime}$}}}}
\put(451,434){\makebox(0,0)[lb]{\smash{{\SetFigFontNFSS{10}{12.0}{\familydefault}{\mddefault}{\updefault}\emph{a)}}}}}
\put(8221,404){\makebox(0,0)[lb]{\smash{{\SetFigFontNFSS{10}{12.0}{\familydefault}{\mddefault}{\updefault}\emph{b)}}}}}
\put(1546,-3811){\makebox(0,0)[lb]{\smash{{\SetFigFontNFSS{10}{12.0}{\familydefault}{\mddefault}{\updefault}$\vec\vartheta = 0$}}}}
\put(9331,-3811){\makebox(0,0)[lb]{\smash{{\SetFigFontNFSS{10}{12.0}{\familydefault}{\mddefault}{\updefault}$\vec\vartheta \not= 0$}}}}
\put(10606,-983){\makebox(0,0)[lb]{\smash{{\SetFigFontNFSS{10}{12.0}{\familydefault}{\mddefault}{\updefault}$V_F, V_H$}}}}
\put(2731,-970){\makebox(0,0)[lb]{\smash{{\SetFigFontNFSS{10}{12.0}{\familydefault}{\mddefault}{\updefault}$V_F, V_H$}}}}
\end{picture}%
\end{center}
\caption{Quadratic coupling of untwisted, $a)$, and twisted, $b)$, open string states to closed
string fluxes. }
\label{fig:flux}
\end{figure}
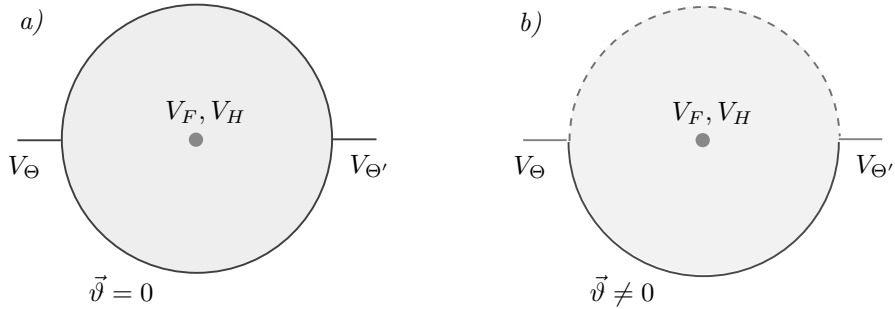

In this section, we would like to present the basic ideas of the general evaluation, using world-sheet methods, of the interactions between
closed string background fluxes and massless open string excitations living on a generic D-brane intersection.
We focus on fermionic terms (like for example mass terms for gauginos),
but our conformal field theory techniques could be applied to other terms of the brane
effective action.
We refer to \cite{Billo':2008sp} for the technical details.
At the lowest order, the fermionic interaction terms can be derived from disk 3-point
correlators involving two vertices describing massless open string fermions and one closed string
vertex describing the background flux, as represented in Fig. \ref{fig:flux}.
At a brane intersection, massless open string modes can arise either from open strings
starting and ending on the same stack of D-branes, or from open strings connecting
two different sets of branes. In the former case the open string fields satisfy
the standard untwisted boundary conditions, and the corresponding vertex operators transform in the
adjoint representation of the gauge group. In the latter case, the string coordinates
satisfy twisted boundary conditions characterized by twist parameters $\vartheta$, and the
associated vertices carry Chan-Paton factors in the bi-fundamental representation of the gauge
group; by inserting twisted open string vertices, one splits the disk boundary into different portions
distinguished by their boundary conditions and Chan-Paton labels, see Fig. \ref{fig:flux}\emph{b)}. We now give some details on
these boundary conditions and later describe the physical vertex operators and their interactions with
RR and NSNS background fluxes.

\paragraph{Boundary conditions and reflection matrices}
\label{sec:bc}
The boundary conditions for the bosonic coordinates $x^M$ ($M=0,\ldots, 9$) of the open string
are given by
\begin{equation}
\Big(\delta_{MN}\,\partial_\sigma x^N+{\rm i}({\cal F}_\sigma)_{MN}\,\partial_\tau
x^N\Big)\Big|_{\sigma=0,\pi}=0~,
\label{bc2}
\end{equation}
where $\delta_{MN}$ is the flat background metric%
\footnote{Here, for convenience, we assume the space-time to have an Euclidean signature. 
} and
\begin{equation}
({\mathcal F}_\sigma)_{MN}
=B_{MN} + 2\pi\alpha' \,({F}_\sigma)_{MN}
\label{calF}
\end{equation}
with $B_{MN}$ the anti-symmetric tensor of the NSNS sector and $({F}_\sigma)_{MN}$
the background gauge field strength at the string end-points $\sigma=0,\pi$.
Introducing the complex variable $z={\rm e}^{\tau+{\rm i}\sigma}$ and the reflection matrices
\begin{equation}
{R}_\sigma=\big(1-{\mathcal F}_\sigma\big)^{-1}\,\big(1+{\mathcal F}_\sigma\big)~,
\label{Rot}
\end{equation}
we can define a monodromy matrix $R \equiv {R}_\pi^{-1} {R}_0$ acting on multi-valued holomorphic fields.
For simplicity, we will take  ${R}_0$ and ${R}_\pi$
to be commuting. Then, with a suitable $\mathrm{SO}(10)$ transformation, 
we can simultaneously diagonalize both matrices and write
\begin{eqnarray}
{R}_\sigma &= \mathrm{diag}\left(\ee^{2\pi\ii\theta_\sigma^1},\ee^{-2\pi\ii\theta_\sigma^1}\,
\ldots,\ee^{2\pi\ii\theta_\sigma^5},\ee^{-2\pi\ii\theta_\sigma^5}\right)~,
\label{diagonalRs}\\
{R}   &= \mathrm{diag}\left(\ee^{2\pi\ii\vartheta^1},\ee^{-2\pi\ii\vartheta^1},\ldots,
\ee^{2\pi\ii\vartheta^5},\ee^{-2\pi\ii\vartheta^5}\right)~,
\label{diagonalR1}
\end{eqnarray}
with $\vartheta^I=\theta_0^I-\theta_\pi^I$.
In this basis, the oscillators corresponding to the complex coordinates act on a twisted vacuum $|\vec{\vartheta}\rangle$
created by the twist operator $\sigma_{\vec{\vartheta}}(z)$.

In the same way, it is necessary to consider also the boundary conditions on the
fermionic fields. Let us only point out that if $\vartheta^I = 0$ there are zero-modes
in the R sector, while if $\vartheta^I = \frac{1}{2}$ there are zero-modes in the NS sector. In these cases the corresponding fermionic vacuum is degenerate and carries the spinor representation of the rotation group acting on the directions in
which the $\vartheta$'s vanish. For this reason 
it is in general necessary to determine the boundary
reflection matrices also in the spinor representation, denoted by $\mathcal{R}_\sigma$. We do not give their explicit expression here, we just want to stress that they depend on the $\theta_\sigma^I$'s and the $\mathrm{SO}(10)$ $\Gamma$-matrices.

Different physical situations can therefore be described through a generic brane intersection,  depending
on the values of the five twists $\vartheta^I$.
When $\vartheta^I=0$ for all $I$'s, all fields are untwisted:
this is the case of the open strings
starting and ending on the same stack of D-branes, which account for dynamical gauge excitations
in the adjoint representation when the branes are space-filling, or for neutral instanton moduli
when the branes are instantonic.
When $\vartheta^4=\vartheta^5=0$ but the $\vartheta^i$'s with
$i=1,2,3$ are non vanishing, only the string coordinates in the
space-time directions are untwisted and describe open strings
stretching between different stacks of D-branes. The corresponding
excitations organize in multiplets that transform in the
bi-fundamental representation of the gauge group and always contain
massless chiral fermions. When suitable relations among the
non-vanishing twists are satisfied ({\it e.g.}
$\vartheta^1+\vartheta^2+\vartheta^3=2$), also massless scalars
appear in the spectrum and they can be combined with the fermions to
form $\mathcal{N}=1$ chiral multiplets suitable to describe the
matter content of brane-world models.
Finally, when $\vartheta^4=\vartheta^5=\frac{1}{2}$, the string
coordinates have mixed Neumann-Dirichlet boundary conditions in the
last four directions and correspond to open strings connecting a
space-filling D-brane  with an instantonic brane. In this situation,
if the $\vartheta^i$'s ($i=1,2,3$) are vanishing, the instantonic
brane describes an ordinary gauge instanton configuration and the
twisted open strings account for the charged instanton moduli of the Atiyah-Drinfeld-Hitchin-Manin 
(ADHM) construction; if
instead also the $\vartheta^i$'s are non vanishing, the instantonic
branes represent exotic instantons of truly stringy nature.
{F}rom these
considerations it is clear that, by considering open strings that are
generically twisted, we can simultaneously treat all interesting configurations.

The reflection matrices we have introduced, directly enter in string amplitudes.
Let us consider the closed string vertex operators corresponding to background fluxes.
In the closed string sector all fields (both bosonic and fermionic) are untwisted
due to the periodic boundary conditions%
\footnote{Even if later we will consider an orbifold compactification, we will
include background fluxes from the untwisted closed string sector only.}. However,
in presence of D-branes, a suitable identification
between the left and the right moving components of the closed string has to be
enforced at the boundary. A non-trivial dependence on the angles $\theta_\sigma^I$ appears through
the matrices $R_\sigma$ or $\mathcal{R}_\sigma$.
For instance, let us describe the RR sector of the Type IIB theory,
where the physical vertex operators for the field strengths of the
anti-symmetric tensor fields are, in the $(-\frac{1}{2},-\frac{1}{2})$ superghost picture,
\begin{equation}
V_{F}(z,\overline z)=\mathcal{N}_{F}\,F_{\mathcal{AB}}~
\ee^{-\ii\pi\alpha'k_{\mathrm L}\cdot k_{\mathrm R}} \big[s_{
\vec{\epsilon}_{\mathcal{A}}} \,\ee^{-\frac{1}{2}\phi} \,\ee^{\ii
\,k_{\mathrm L} \cdot X}\big](z)
 \times
\big[\widetilde s_{ \vec{\epsilon}_{\mathcal{B}}}
 \,\ee^{-\frac{1}{2}\widetilde\phi} \,\ee^{\ii \,k_{\mathrm R} \cdot \widetilde X}\big](\overline z)~.
\label{vertexRR}
\end{equation}
In this expression the
index $\mathcal{A}=1,\ldots,16$ labels a spinor representation of $\mathrm{SO}(10)$ with definite
chirality and runs over all possible choices of signs in the weight vector
$\vec{\epsilon}_{\mathcal{A}} = \frac{1}{2}\Big(\pm,\pm,\pm,\pm,\pm\Big) .
\label{epsilonA}$
The symbol $s_{\vec{q}}(z)$ stands for the fermionic
spin field
$s_{\vec{q}}(z)=\ee^{\ii\sum_Iq^I\varphi^I(z)},
\label{spinfield}$
where $\varphi^I(z)$ are the fields that bosonize the world-sheet fermions according to
$\Psi^I=\ee^{\ii\varphi^I}$ (up to cocycle factors). $\phi(z)$ is the boson entering the superghost
fermionization formulas. Finally, 
$\mathcal{N}_F$ is a normalization factor,
$k_{\mathrm L}$ and $k_{\mathrm R}$ are the left and right momenta, and the
tilde sign denotes the right-moving components.
Furthermore, the factor $\ee^{-\ii\pi\alpha'k_{\mathrm L}\cdot k_{\mathrm R}}$ is a cocycle that allows
for an off-shell extension of the closed string vertex with $k_{\mathrm L}\not =k_{\mathrm R}$, as discussed in Ref. \cite{Bertolini:2005qh}. The bi-spinor polarization $F_{\mathcal{AB}}$
comprises all RR field strengths
of the Type IIB theory according to
\begin{equation}
F_{\mathcal{AB}}=\sum_{n=1,3,5}\frac{1}{n!}\,F_{M_1\ldots M_n}
\left(\Gamma^{M_1\ldots M_n}\right)_{\mathcal{AB}}~,
\label{F135}
\end{equation}
even if in our applications only the three-form part will play a role.
In the presence of D-branes the left and right
moving components of the vertex operator $V_F$ must be identified using the reflection rules
discussed above. In practice (see for example Ref. \cite{Bertolini:2005qh} for more details)
this amounts to set
\begin{equation}
\widetilde X^M(\overline{z}) = (R_0)^M_{~N}\,X^N(\overline{z})
\quad,\quad \widetilde s_{ \vec{\epsilon}_{\mathcal{A}}}(\overline{z})=
(\mathcal{R}_0)^{\mathcal{A}}_{~\,\mathcal B}\,
s_{ \vec{\epsilon}_{\mathcal{B}}}(\overline{z})\quad,\quad\widetilde\phi(\overline{z})=
\phi(\overline{z})
\label{leftright}
\end{equation}
and modify the cocycle factor in the vertex operator (\ref{vertexRR}) to
$\ee^{-\ii\pi\alpha'k_{\mathrm L}\cdot (k_{\mathrm R}R_0)}$.
As a consequence of the identifications (\ref{leftright}),
the RR field-strength $F_{\mathcal{AB}}$
gets replaced by the bi-spinor polarization $(F\,\mathcal{R}_0)_{\mathcal{AB}}$ that
incorporates also the information on the type of boundary conditions enforced by the D-branes.

The same happens for the NSNS sector of the closed string. Here it is possible to write
an effective BRST invariant vertex operator for the derivatives of the anti-symmetric tensor $B$
that are related to the three-form flux $H$.
In the $(0,-1)$ superghost picture%
, this effective vertex is
\begin{equation}
V_H(z,\overline{z}) = {\mathcal{N}_H}\,\big(\partial_MB_{NP}\big)\,
\ee^{-\ii\pi\alpha'k_{\mathrm L}\cdot k_{\mathrm R}}
\big[\psi^M\psi^N\ee^{\ii \,k_{\mathrm L} \cdot X}\big](z) \times
\big[\widetilde \psi^P\,\ee^{-\widetilde\phi} \,\ee^{\ii
\,k_{\mathrm R} \cdot \widetilde X}\big] (\overline z)~,
\label{vertexNS}
\end{equation}
where again we have introduced a normalization factor and a cocycle. When we insert this vertex in
a disk diagram, we must identify the left and right moving sectors using the reflection rules
(\ref{leftright}) supplemented by
\begin{equation}
\widetilde \psi^M(\overline{z}) = (R_0)^M_{~N}\psi^N(\overline{z})~.
\label{leftright1}
\end{equation}
Consequently, in (\ref{vertexNS}) the polarization $(\partial B)$ is effectively replaced by
$(\partial BR_0)$. Notice that the NSNS polarization combines with the boundary reflection matrix in the
vector representation $R_0$, in contrast to the RR case
where one finds instead the reflection matrix in the spinor representation $\mathcal{R}_0$.

We have now the ingredients to evaluate the string correlation functions among two massless open string fermions $\Theta_{\mathcal{A}}$, described by the vertex operator 
\begin{equation}
V_{\Theta}(z) = \mathcal{N}_{\Theta}\,\Theta_{\!\mathcal{A}}
\big[\sigma_{\vec{\vartheta}}\,s_{ \vec{\epsilon}_{\mathcal{A}}+\vec{\vartheta}}\,
\ee^{- {\frac{1}{2}\phi}} \,\ee^{\ii \,k \cdot X}\big](z)
\label{vertexferm}
\end{equation}
where $\sigma_{\vec{\vartheta}}(z)$ is the bosonic twist field, and the
background closed string flux, as represented in Fig. \ref{fig:flux}.

Let us analyze first the interaction with the RR flux.
It is given by the amplitude 
\begin{equation}
\mathcal{A}_{F}= \Big\langle
V_{\Theta}(x)\,V_F(z,\overline{z})\,V_{\Theta'}(y) \Big\rangle
= c_F~ {\Theta}_{\!\mathcal {A}_1}(F\mathcal{R}_0)_{\mathcal{A}_2\mathcal{A}_3}\,
{\Theta'}_{\!\!\mathcal{A}_4}\,\times\,
A^{\mathcal{A}_1\mathcal{A}_2\mathcal{A}_3\mathcal{A}_4}
\label{amplF}
\end{equation}
where the prefactor $c_F$
 accounts for the normalizations of the vertex operators and the topological
normalization $\mathcal{C}_{(p+1)}$
of any disk amplitude with the boundary conditions of a D$p$-brane.
The last factor in (\ref{amplF}) is the four-point correlator.
Collecting all contributions, the result for the amplitude is
\begin{equation}
\mathcal{A}_{F}= -8c_F\Theta'\Gamma^M\Theta\,\big[F\mathcal{R}_0(2I_{1}-I_2)\big]_M
+\frac{4c_F}{3!}\Theta'\Gamma^{MNP}\Theta\,\big[F\mathcal{R}_0I_2\big]_{MNP}~,
\label{amplFfinal}
\end{equation}
where $I_1$ and $I_2$ are two $\vec\vartheta$-dependent diagonal matrices.

Similarly, the fermionic couplings induced by the NSNS three-form flux  arise from
the following mixed disk amplitude
\begin{equation}
\mathcal{A}_{H}= \Big\langle
V_{\Theta}(x)\,V_H(z,\overline{z})\,V_{\Theta'}(y) \Big\rangle
= {c_H}~ {\Theta}_{\!\mathcal A}(\partial B{R}_0)_{{MNP}}\,
{\Theta'}_{\!\mathcal B}\,\times\,
A^{\mathcal{AB};MNP}~.
\label{amplH}
\end{equation}
The result is the NSNS counterpart of the RR amplitude (\ref{amplFfinal}) on a generic D-brane intersection,
and shares with it the same type of fermionic structures:
\beqa
{}~\mathcal{A}_{H} &=& -{4c_H}\Theta'\Gamma^N\Theta \,\delta^{MP}\,\big[\partial BR_0(2I_{1}-I_2)\big]_{[MN]P}
+{2c_H}\Theta'\Gamma^{MNP}\Theta\,\big[\partial B R_0I_2\big]_{MNP}.
\label{amplHfinal}
\eeqa

\paragraph{Flux couplings with untwisted open strings ($\vec{\vartheta}=0$)}
\label{sec:effects}
We now exploit these results  to analyze how constant background fluxes
couple to untwisted open strings, {\it i.e.} strings starting and ending on a single stack of D-branes.
This corresponds to set $\vec{\vartheta}=0$ in all previous formulas, which drastically simplify.
Note that the condition $\vec{\vartheta}=0$ implies that $\vec{\theta}_0=\vec{\theta}_\pi$, so that the
reflection rules are the same at the two string end-points. We can
distinguish two cases, namely when these reflection rules are just signs
({\it i.e.} $\theta_\sigma^I=0$ or $1$), and when they instead depend on generic angles $\theta_\sigma^I$.
In the first case, which will be considered in the following, the branes are unmagnetized,  while the second corresponds to
magnetized branes.

Since we are interested in constant background fluxes, we can set the momentum of the closed string
vertices to zero. 
Using this result in the RR and NSNS amplitudes (\ref{amplFfinal}) and (\ref{amplHfinal}),
the total flux amplitude is
\begin{equation}
\mathcal A \equiv \mathcal{A}_F + \mathcal{A}_H =-2\pi\ii\,
\Theta\Gamma^{MNP}\Theta\,\Big[\frac{c_F}{3}\big(F\mathcal{R}_0\big)_{MNP}
+c_H\big(\partial B R_0\big)_{MNP}\Big]~. \label{ampltot}
\end{equation}
Here we used the fact that the untwisted fermions $\Theta$ and
$\Theta'$ in (\ref{amplFfinal}) and (\ref{amplHfinal}) actually
describe the same field and only differ because they carry opposite
momentum. For this reason we multiplied the above amplitudes by a
symmetry factor of 1/2 and dropped the $'$ without introducing
ambiguities.

It is clear from Eq. (\ref{ampltot}) that once the flux configuration is given, the structure of the
fermionic couplings for different types of D-branes depends crucially
on the boundary reflection matrices $R_0$ and $\mathcal R_0$.
Notice that the RR piece of the amplitude (\ref{ampltot}) is generically non zero for 1-form, 3-form
and 5-form fluxes. However, from now on we will restrict our analysis only to the 3-form  and hence the
bi-spinor to be used is simply
\begin{equation}
F_{\mathcal{AB}} = \frac{1}{3!} F_{MNP}
\left(\Gamma^{MNP}\right)_{\mathcal{AB}}~.
\label{F3}
\end{equation}

The normalization factors $c_F$ and $c_H$ are determined to be
\begin{equation}
c_H=c_F/g_s
\label{cfch}
\end{equation}
by quadratic terms of the bulk theory in the ten-dimensional Einstein frame. 
Taking all this into account, we can rewrite the total amplitude (\ref{ampltot}) as
\begin{equation}
\mathcal A \equiv \mathcal{A}_F + \mathcal{A}_H =-\frac{2\pi\ii}{3!}\,c_F\,
\Theta\Gamma^{MNP}\Theta\,
T_{MNP}
\label{ampltot1}
\end{equation}
where
\begin{equation}
T_{MNP} = \big(F\mathcal{R}_0\big)_{MNP}+\frac{3}{g_s}\,
\big( \partial B R_0\big)_{[MNP]} ~.
\label{tmnl}
\end{equation}

Up to now we have used a ten-dimensional notation. However, since we are interested
in studying the flux induced couplings for gauge theories and instantons in four dimensions, it becomes
necessary to split the indices $M,N,\ldots=0,1,\ldots,9$ appearing in the above equations into four-dimensional space-time indices $\mu,\nu,\ldots=0,1,2,3$, and six-dimensional indices
$m,n,\ldots=4,5,\ldots,9$ labeling the directions of the internal space.
In the following, we will consider only internal fluxes,
like $F_{mnp}$ or $(\partial B)_{mnp}$, which preserve the four-dimensional Lorentz invariance, and
the fermionic amplitudes we will compute are of the form
\begin{equation}
\mathcal A =-\frac{2\pi\ii}{3!}\,c_F\,\Theta\Gamma^{mnp}\Theta\,T_{mnp}~ .
\label{ampltot2}
\end{equation}

To more clearly expose  the structure of the flux-induced
fermionic masses,  the fermion bilinear $\Theta\Gamma^{mnp}\Theta$
can be written using a four-dimensional spinor notation.  
One can show that
\begin{equation}
\Theta \Gamma^{mnp}\Theta\,T_{mnp} = -\ii\,\Theta^{\alpha A}
\Theta_{\alpha}^{{\phantom\alpha} B}
\big(\overline\Sigma^{mnp}\big)_{AB}\,T_{mnp}^{\mathrm{IASD}}
-\ii\,\Theta_{\dot\alpha A} \Theta^{\dot\alpha}_{{\phantom\alpha}
B}\big(\Sigma^{mnp}\big)^{AB} \,T_{mnp}^{\mathrm{ISD}}
\label{decomp1}
\end{equation}
where $\alpha$ ($\dot\alpha$) are chiral (anti-chiral) indices in four dimensions,
and the lower (upper) indices $A$ are chiral (anti-chiral) spinor indices of
the internal six dimensional space. 
$\Sigma^{mnp}$ and $\overline\Sigma^{mnp}$ are respectively the chiral and anti-chiral blocks
of $\gamma^{mnp}$ and
\begin{equation}
T_{mnp}^{\mathrm{ISD}} =\frac{1}{2}\big(T-\ii*_6\!T\big)_{mnp}\quad,\quad
T_{mnp}^{\mathrm{IASD}} =\frac{1}{2}\big(T+\ii*_6\!T\big)_{mnp}~.
\label{isdiasd}
\end{equation}
Fixing a complex structure, the three-form tensors
$T^{\mathrm{ISD}},T^{\mathrm{IASD}}$  can be decomposed into their
(3,0),(2,1),(1,2) and (0,3) parts.

Let us now use this information to rewrite the fermionic terms we have previously discussed, focusing in particular on space-filling D3- and D(-1)-branes (or D-instantons)
on flat space.
In the case of D3-branes, we can use a Minkowski
signature and the Majorana-Weyl fermion $\Theta$ decomposes in four-dimensional chiral and anti-chiral
components, which are related by charge conjugation and assembled into four Majorana spinors.
These are the four gauginos living on the world-volume of the D3-brane, and for future
notational convenience we will denote their chiral and anti-chiral parts
as $\Lambda^{\alpha A}$ and $\bar\Lambda_{\dot\alpha A}$ (instead of $\Theta^{\alpha A}$
and $\Theta_{\dot\alpha A}$).
Inserting the specific contribution of the reflection matrices, the general
expression (\ref{ampltot2}) becomes
\begin{equation}
\mathcal A_{\mathrm D3} =
\frac{2\pi\ii}{3!}\,c_F(\Lambda)\,{\mathrm{Tr}}\Big[\, \Lambda^{\alpha A}
\Lambda_{\alpha}^{{\phantom \alpha}B}
\big(\overline\Sigma^{mnp}\big)_{AB}\,G_{mnp}^{\mathrm{IASD}} -
\bar\Lambda_{\dot\alpha A}\bar\Lambda^{\dot\alpha}_{{\phantom
\alpha}B}
\big(\Sigma^{mnp}\big)^{AB}\,\big(G_{mnp}^{\mathrm{IASD}}\big)^*
\,\Big] \label{massD3}
\end{equation}
where we have made explicit the colour trace
generators 
and  has been possible to write
\begin{equation}
T_{mnp}=(*_6 F)_{mnp} - \frac{1}{g_s}\,H_{mnp} =  \re\big( \!*_6 \!G-\ii G\big)_{mnp}~,
\label{tD31}
\end{equation}
by introducing\footnote{Self-duality of type IIB can be used to
promote this expression to its $SL(2,\mathbb{Z})$-covariant version $G=F-\tau H$ with
$\tau=C_0-\ii \ee^{-\varphi}$. A direct evaluation of the $C_0$-dependent term however requires
a string amplitude involving two closed and two open string insertions in the disk. }
\begin{equation}
G=F-\frac{\ii}{g_s}\,H ~.
\label{G}
\end{equation}

{F}rom the explicit expression of the amplitude (\ref{massD3})
we see that an IASD $G$-flux configuration
induces a Majorana mass  %
for the
gauginos leading to supersymmetry breaking on the gauge theory.
 Notice
that the mass terms for the two different chiralities are complex
conjugate of each other: $T^{\mathrm{IASD}}=-\ii\,
G^{\mathrm{IASD}}$ and $T^{\mathrm{ISD}}=\ii(G^{\mathrm{IASD}})^*$. This
is a consequence of the Majorana condition that the four-dimensional
spinors inherit from the Majorana-Weyl condition of the fermions in
the original ten-dimensional theory.

If we decompose $G^{\mathrm{IASD}}$, we see  that a $G$-flux of type $(1,2)_{\mathrm P}$
gives mass to the three gauginos transforming non-trivially under
$\mathrm{SU}(3)$ but keeps the $\mathrm{SU}(3)$-singlet gaugino massless, thus
preserving ${\mathcal N}=1$ supersymmetry. On the other hand, a $G$-flux of type $(3,0)$,
or $(2,1)_{\mathrm{NP}}$ gives mass also to the $\mathrm{SU}(3)$-singlet gaugino.

Things are rather different instead on D-instantons. Indeed, we have
\begin{equation}
\mathcal A_{{\mathrm D}(-1)} =
\frac{2\pi\ii}{3!}\,c_F(\Theta)\,\Big[\, \Theta^{\alpha A}
\Theta_{\alpha}^{{\phantom\alpha} B}
\big(\overline\Sigma^{mnp}\big)_{AB}\,G_{mnp}^{\mathrm{IASD}} +\bar
\Theta_{\dot\alpha A}\bar \Theta^{\dot\alpha}_{{\phantom \alpha}B}
\big(\Sigma^{mnp}\big)^{AB}\,G_{mnp}^{\mathrm{ISD}} \,\Big]
\label{massD-1}
\end{equation}
where now the prefactor $c_F(\Theta)$ contains the topological
normalization of the D$(-1)$ disks.
{}From the amplitude (\ref{massD-1}) we explicitly see that
both IASD and ISD components of the $G$-flux couple to the D-instanton fermions;
however, the couplings are different and independent for the two chiralities since
they are not related by complex conjugation, as always in Euclidean spaces.
In particular, comparing Eqs. (\ref{massD3}) and
(\ref{massD-1}), we see that an ISD $G$-flux does not give a mass to any gaugino
but instead induces a ``mass'' term for the anti-chiral instanton zero-modes which are therefore
lifted. This effect plays a crucial role in discussing the non-perturbative contributions
of the exotic D-instantons for which the neutral anti-chiral zero modes $\bar \Theta_{\dot\alpha A}$ must be removed or lifted by some mechanism.

Let us mention that these results can be generalized in a rather straightforward way
to branes with a non-trivial magnetization on their world-volume.

\paragraph{Flux couplings with twisted open strings ($\vec{\vartheta}\not =0$)}
\label{sec:twisted}

As we have emphasized, the general world-sheet calculation allows to obtain the couplings between
closed string fluxes and open string fermions at a generic D-brane
intersection, even for non-vanishing twist parameters
$\vec{\vartheta}$. 
Here we just present the result for  a simple case
of such twisted amplitudes, which will be relevant later.

The case we discuss is that of the three-form flux couplings with the
twisted fermions stretching between a D3-brane and a D-instanton,
which represent the charged (or flavored) fermionic moduli of the
${\cal N}=4$ ADHM construction of instantons, and are usually denoted as $\mu^A$
and $\bar\mu^A$ depending on the orientation.

Collecting the contributions coming from the correlator, the result is
\begin{equation}
{\mathcal A}_{\mathrm{D3/D(-1)}} \equiv {\mathcal A}_F +{\mathcal
A}_H =\frac{4\pi\ii}{3!}\,c_F(\mu)\,{\bar\mu}^{A}\mu^{B}
\,\big(\overline\Sigma^{mnp}\big)_{AB}\,G_{mnp}^{\mathrm{IASD}}~.
\label{mumubtot}
\end{equation}
This amplitude, together with the one
in \eq{massD-1}, accounts for the flux induced fermionic couplings on
the D-instanton effective action.

\section{Instantons in gauge and string theory}
\label{ginst}

We now want to review some notions on instantons, as the interplay between them and background fluxes is very deep.

Instantons in gauge theories, defined in Minkowski spacetime, describe tunneling processes from one vacuum to another. The simplest models which exhibit this phenomenon are the quantum mechanical point particle with a double-well potential having two vacua, or a periodic potential with infinitely many vacua. There is no classical allowed trajectory for a particle to travel from one vacuum to the other, but quantum mechanically tunneling occurs. In the Euclidean space-time, where path integrals are more conveniently computed, instantons are defined as finite action solutions to the field equations of motion.  For instance, in the four-dimensional SU($N$) pure Yang-Mills, this implies that the gauge field strenght obeys the (anti)self-duality condition
\beq
\label{eqinst}
F = \pm {}*F~.
\eeq
The most powerful method to solve the (anti)self-dual equations (\ref{eqinst}) is the ADHM construction \cite{adhm}. 
It realizes the instanton moduli space as a hyper-K\"{a}hler quotient of a flat space by an auxiliary U($k$) gauge theory.
The Higgs branch of this U($k$) theory, which is related to the moduli space (which is the space of inequivalent solutions of self-dual SU($N$) Yang-Mills equations),  is defined through a triplet of algebraic equations, the ADHM constraints, for each solution of which a solution to the set of equations (\ref{eqinst}) can be built.
Euclidean path integral requires to keep in consideration all these configurations, where fields assume a non-trivial profile, by summing over them.
The contribution of instantons to the path integral is exponentially suppressed. Moreover, when fermions are present, strong selection rules appear and may eventually lead to a vanishing instanton contribution.
The partition function is obtained by integrating over all the possible inequivalent histories, \emph{i.e.} over the inequivalent configurations of the fields. 
This can be traded for an integral over the moduli space.

Instantons play a leading role in the understanding of non-perturbative regime of four-dimensional supersymmetric gauge theories. As shown by Affleck, Dine and Seiberg \cite{Affleck:1983mk}, instantons in SQCD with gauge group SU($N_c)$ and $N_f$ massless flavors generate a superpotential in the case $N_f = N_c -1$.
Even in cases where instantons do not generate such superpotential, they can deform the complex structure of the moduli space of supersymmetric vacua \emph{via} the creation of an F-term \cite{Beasley:2004ys}, which cannot be integrated to retrieve a corresponding superpotential but is nevertheless a genuine F-term.

It was pointed out in 1995 \cite{Witten:1995gx} that gauge instantons could have a realization in the frame of string theory. Systems of suitably chosen D-branes, D-instantons and Euclidean branes can indeed support the stringy description of gauge instantons, naturally embedding the ADHM construction.
It was then argued that non-perturbative effects, such as superpotentials arising from instantons and gaugino condensation, could for instance solve the problem of moduli stabilization. Moreover it was found that string theory could provide new kinds of instantons, called exotic, whose field theory explanation is beginning recently to be uncovered \cite{Amariti:2008xu,exoticbis,Billo:2009di}.
The study of these exotic instanton configurations has led to
interesting results in relation to moduli stabilization, (partial)
supersymmetry breaking and even fermion masses and Yukawa couplings. A
delicate point about these stringy instantons concerns the presence
of neutral anti-chiral fermionic zero-modes which completely
decouple from all other instanton moduli, contrarily to what happens
for the usual gauge theory instantons where they act as Lagrange
multipliers for the fermionic ADHM constraints.
In order to get non-vanishing contributions to the effective action
from such exotic instantons, it is therefore necessary to remove
these anti-chiral zero modes 
or lift them by some mechanism. The
presence of internal background fluxes allows for such a lifting
and points to the existence of an intriguing interplay among soft
supersymmetry breaking, moduli stabilization, instantons and
more-generally non-perturbative effects in the low-energy theory
which may lead to interesting developments and applications.

\paragraph{The D3/D(--1)-branes on $\mathbb{C}^3/\big(\mathbb{Z}_2\times \mathbb{Z}_2\big)$}
\label{d3d01}

\begin{figure}[hbt]
\begin{center}
\begin{picture}(0,0)%
\includegraphics{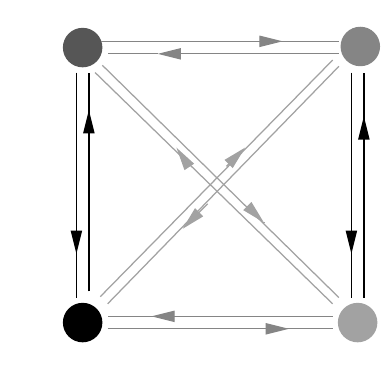}%
\end{picture}%
\setlength{\unitlength}{1381sp}%
\begingroup\makeatletter\ifx\SetFigFontNFSS\undefined%
\gdef\SetFigFontNFSS#1#2#3#4#5{%
  \reset@font\fontsize{#1}{#2pt}%
  \fontfamily{#3}\fontseries{#4}\fontshape{#5}%
  \selectfont}%
\fi\endgroup%
\begin{picture}(4559,4450)(211,-4247)
\put(4726,-136){\makebox(0,0)[lb]{\smash{{\SetFigFontNFSS{8}{9.6}{\familydefault}{\mddefault}{\updefault}$N_1$}}}}
\put(4726,-4111){\makebox(0,0)[lb]{\smash{{\SetFigFontNFSS{8}{9.6}{\familydefault}{\mddefault}{\updefault}$N_2$}}}}
\put(301,-136){\makebox(0,0)[lb]{\smash{{\SetFigFontNFSS{8}{9.6}{\familydefault}{\mddefault}{\updefault}$N_0$}}}}
\put(226,-3961){\makebox(0,0)[lb]{\smash{{\SetFigFontNFSS{8}{9.6}{\familydefault}{\mddefault}{\updefault}$N_3$}}}}
\end{picture}%
\end{center}
\caption{The quiver diagram encoding the field content and the charges for
fractional D-branes of the orbifold
$\mathbb{C}^3/(\mathbb{Z}_2\times\mathbb{Z}_2)$. The dots represent the branes associated with the irrep $R_A$
of the orbifold group. A stack of $N_A$ such branes supports a $\mathrm{U}(N_A)$ gauge theory.
An oriented link from the $A$-th to the $B$-th dot
corresponds to a chiral multiplet transforming in the
$\big({N_A},{\overline N_B}\big)$ representation of the gauge group and in the $R_A \otimes R_B$ representation of the orbifold group.}
\label{fig:quiver}
\end{figure}

We now focus our attention on a particular instanton configuration. We indeed consider  $N$ parallel D3-branes and $k$ D(-1)-branes on the orbifold $\mathbb{C}^3/\big(\mathbb{Z}_2\times \mathbb{Z}_2\big)$ (see Fig. \ref{fig:quiver}).
The fundamental types of D-branes which can be placed transversely
to an orbifold space are called fractional branes. Such branes must be localized at one of the
fixed points of the orbifold group. For
simplicity we focus on fractional D3-branes sitting at a specific
fixed point (say, the origin) and work around this configuration.
When the D-branes are placed in the $\mathbb{C}^3/\big(\mathbb{Z}_2\times \mathbb{Z}_2\big)$ 
orbifold, the supersymmetry of the gauge theory is reduced to $\mathcal N=1$ and only
the $\big(\mathbb{Z}_2 \times \mathbb{Z}_2\big)$-invariant components of  the fields 
are retained. 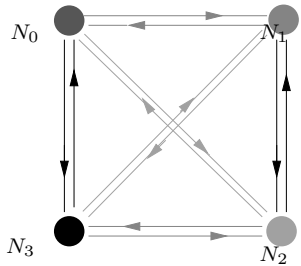

In the unresolved (singular) orbifold limit, which from the string point of view corresponds
to switching off the fluctuations of all twisted closed string fields, we have that the gauge kinetic functions $\tau_A$ ($A = 0, \ldots 3 $) are the same 
for all $A$'s. However, by turning on twisted closed string moduli, one can introduce
differences among the $\tau_A$'s and thus distinguish the gauge couplings of
the various group factors.

Gauge instantons correspond to D$(-1)$-branes that sit on non-empty
nodes of the quiver diagram, so that their number $k_A$ can be interpreted as the second Chern class
of the Yang-Mills bundle of the $\mathrm{U}(N_A)$ component of the gauge group.
Stringy instantons correspond instead to D$(-1)$-branes
occupying empty nodes of the quiver. 
The results obtained in Section \ref{sec:CFT}  extend straightforwardly to these less
supersymmetric theories. In particular for pure ${\cal N}=1$ SYM,
the flux couplings for both gauge and exotic instantons follow from
previous formulas by restricting the
spinor components to $A=B=0$. 
 The flux induced terms in the instanton moduli action are thus%
\begin{equation}
S_{\mathrm{inst}}^{\mathrm{flux}} = - \mathcal A^{\rm flux}_{{\mathrm D}(-1)} - \mathcal A^{\rm flux}_{\mathrm{D3/D(-1)}}
\label{sflux}
\end{equation}
where $\mathcal A^{\rm flux}_{{\mathrm D}(-1)}$ and $\mathcal A^{\rm flux}_{\mathrm{D3/D(-1)}} $
are the $A=B=0$ parts of the amplitudes (\ref{massD-1})
and (\ref{mumubtot}), {\it i.e.}
\begin{equation}
\begin{aligned}
\mathcal A^{\mathrm{flux}}_{{\mathrm D}(-1)}{\phantom{\vdots}} &= -2\pi\ii\,c_F(\theta)\,\theta^\alpha\theta_\alpha\,{G}_{(3,0)}
+ 2\pi\ii\,c_F(\lambda)\,\lambda_{\dot\alpha}\lambda^{\dot\alpha}\,{G}_{(0,3)}~,\\
\mathcal A^{\mathrm{flux}}_{\mathrm{D3/D(-1)}}{\phantom{\vdots}}
&=-4\pi\ii\,c_F(\mu)\,\bar\mu\,\mu\,{G}_{(3,0)}~.
\end{aligned}
\label{sflux01}
\end{equation}
where
\begin{equation}
 \Theta^{\alpha 0}\, \sim\, g_0\, \theta^{\alpha}\,\quad\mbox{,}\quad
\Theta_{\dot\alpha 0} \,\sim\, \lambda_{\dot\alpha}\,\quad\mbox{and}\quad\mu^0\,\sim\,g_0\,\mu \quad\mbox{,}\quad\bar \mu^0\,\sim\,g_0\,\bar \mu~.
\label{Thetalambda}
\end{equation}
and $g_0$ is the D(-1) gauge coupling.

The $\theta^2$ term represents the auxiliary component of the gauge
kinetic function $\tau_A=\tau/4$, which is therefore promoted to the
full chiral superfield $\tau_A(\theta)=\tau(\theta)/4$. The other two
terms represent the explicit effects of a
background $G$-flux on the instanton moduli space, and are the
strict analogue for the instanton action of the soft supersymmetry
breaking terms of the gauge theory. In particular, the $\bar\mu\mu$
term is related to the IASD flux component $G_{(3,0)}$ which is
responsible for the gaugino mass $m_\Lambda$, while the $\lambda^2$
term represents a truly stringy effect on the instanton moduli space. Indeed, it can be demonstrated that it is related to the ISD flux component $G_{(0,3)}$ which gives
rise to the gravitino mass $m_{3/2}$.
For the exotic instanton configurations, the neutral
anti-chiral fermionic moduli $\lambda_{\dot\alpha}$ do not couple
to anything, therefore one way to avoid a trivial vanishing result upon integration
over the moduli space is to lift them by coupling
the fractional D-instanton to an ISD $G$-flux of type (0,3).

\section{Non-perturbative flux-induced effective interactions in SQCD-like models}
\label{secn:fluxeffects}

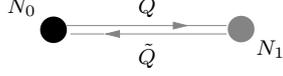
\begin{figure}[hbt]
\begin{center}
\begin{picture}(0,0)%
\includegraphics{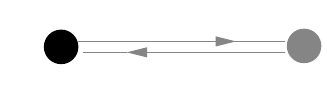}%
\end{picture}%
\setlength{\unitlength}{1381sp}%
\begingroup\makeatletter\ifx\SetFigFontNFSS\undefined%
\gdef\SetFigFontNFSS#1#2#3#4#5{%
  \reset@font\fontsize{#1}{#2pt}%
  \fontfamily{#3}\fontseries{#4}\fontshape{#5}%
  \selectfont}%
\fi\endgroup%
\begin{picture}(4455,1375)(361,-1097)
\put(2701,-61){\makebox(0,0)[lb]{\smash{{\SetFigFontNFSS{8}{9.6}{\familydefault}{\mddefault}{\updefault}$Q$}}}}
\put(2701,-961){\makebox(0,0)[lb]{\smash{{\SetFigFontNFSS{8}{9.6}{\familydefault}{\mddefault}{\updefault}${\tilde Q}$}}}}
\put(4801,-811){\makebox(0,0)[lb]{\smash{{\SetFigFontNFSS{8}{9.6}{\familydefault}{\mddefault}{\updefault}$N_1$}}}}
\put(376,-61){\makebox(0,0)[lb]{\smash{{\SetFigFontNFSS{8}{9.6}{\familydefault}{\mddefault}{\updefault}$N_0$}}}}
\end{picture}%
\end{center}
\caption{This simple quiver gauge theory is (from the point of view of one of the nodes) 
just $\mathcal{N}=1$ SQCD.}
\label{fig:quiver_SQCD}
\end{figure}

We now study the SQCD-like model of Fig. \ref{fig:quiver_SQCD}, realized by taking in the previous quiver gauge theory
\begin{equation}
N_2=N_3=0\quad\mbox{with}\quad N_0~\mbox{and}~N_1~\mbox{arbitrary}~.
\label{NA}
\end{equation}

This brane system describes a $\mathcal N=1$ theory with gauge group 
$\mathrm U(N_0)\times \mathrm U(N_1)$ and a single bifundamental multiplet
\begin{equation}
 \label{Phimult}
\Phi^1(x,\theta)\equiv \Phi(x,\theta)=\phi(x)+\sqrt2 \theta \psi(x)+ \theta^2 F(x)~,
\end{equation}
which in block form is
\begin{equation}
\Phi=\left(
         \begin{array}{cc}
           0 & Q^{u}_{~ f} \\
           \widetilde Q^{f}_{~ u} & 0 \\
         \end{array}
       \right)
\label{Phi}
\end{equation}
with $u=1,\ldots N_0$ and $f=1,\ldots N_1$. The two off-diagonal blocks $Q$ and $\widetilde Q$ represent the quark and anti-quark superfields,
which transform respectively in the fundamental and anti-fundamental of $\mathrm U(N_0)$, and in the anti-fundamental and fundamental of $\mathrm U(N_1)$. Both quarks and anti-quarks are neutral
under the diagonal $\mathrm U(1)$ factor of the gauge group, which decouples. 
On the other hand,
the relative $\mathrm U(1)$ group, under which both $Q$ and $\widetilde Q$ are charged, is IR free 
and thus at low energies the resulting effective gauge group
is $\mathrm {SU}(N_0)\times \mathrm {SU}(N_1)$. Therefore,
from the point of view of, say, the $\mathrm{SU}(N_0)$ factor
this theory is just $\mathcal N=1$ SQCD with $N_c=N_0$ colors
and $N_f=N_1$ flavors. In the following we will study
the non-perturbative properties of this theory in the
Higgs phase where the gauge invariance is completely broken by
giving (large) vacuum expectation values to the lowest components of the matter superfields.
This requires $N_f\geq N_c-1$. The moduli space of this SQCD is obtained by imposing
the D-flatness conditions. Even if the effective gauge
group is $\mathrm{SU}(N_c)$, we have to impose the D-term equations also for the (massive)
$\mathrm{U}(1)$ factors to obtain the correct moduli space of the quiver theory; 
in our case these D-term conditions lead to the constraint
\begin{equation}
Q\,\bar Q - \bar{\widetilde Q}\,\widetilde Q = \xi\,\one_{N_c\times N_c}
\label{D-flat1}
\end{equation}
where $\xi$ is a Fayet-Iliopoulos parameter related to twisted closed string fields which vanish
in the singular orbifold limit. For $N_f\geq N_c$ the D-term constraints allow for flat directions parameterized by meson fields
\begin{equation}
M^{f_1}_{~f_2} \equiv {\widetilde Q}^{f_1}_{~u} \,Q^{u}_{~f_2}
\label{meson}
\end{equation}
and baryon fields
\begin{equation}
B_{f_1\ldots f_{N_c}}=\epsilon_{u_1\ldots u_{N_c}}\, Q^{u_1}_{~f1} \ldots Q^{u_{N_c}}_{~\,f_{N_c}} 
\quad,\quad
{\widetilde B}^{f_1\ldots f_{N_c}}
=\epsilon^{u_1\ldots u_{N_c}}\, {\widetilde Q}^{f1}_{~u_1} \ldots 
{\widetilde Q}^{f_{N_c}}_{~\,u_{N_c}} 
\label{baryon}
\end{equation}
which are subject to constraints whose specific form depends on the difference $(N_f-N_c)$. These are the good observables 
of the low-energy theory in the Higgs phase.
For $N_f=N_c-1$, instead, the baryons cannot be formed and only the meson fields are present.

The non-perturbative effects produced by a configuration of fractional D-instantons 
with numbers $k_A$ can be analyzed by studying the centered partition function
\begin{equation}
{W}_{\mathrm{n.p.}}  = \int d\,{\widehat{\mathfrak M}}\,\,\,\prod_{A=0}^3\!\Big(
M_s^{\,k_A \beta_A}\,\ee^{2\pi \ii k_A \tau_A }\,
\ee^{-\mathrm{Tr}_{k_A}[S]}\Big)~,
\label{Z}
\end{equation}
where the integration is over all moduli except for the
center of mass supercoordinates $x^\mu$ and $\theta^\alpha$. These centered moduli are collectively denoted by $\widehat{\mathfrak M}$.
The action $\mathrm{Tr}_{k_A}[S]$  depends on the entire matter fields and on moduli, where
the latter are restricted to their $\mathbb Z_2 \times \mathbb Z_2$ invariant blocks for each $A$, while
the term $2\pi\ii k_A\tau_A$ represents the classical action of $k_A$ fractional
D-instantons of type $A$. The power of the string scale $M_s$ compensates for the scaling dimensions of 
the measure over the centered moduli space and $\beta_A$ is the one-loop $\beta$-function coefficient of the $\NN = 1$ SU($N_A$) gauge theory with $\sum_{I=1}^3 N_{A\otimes I}$ fundamentals and anti-fundamentals.

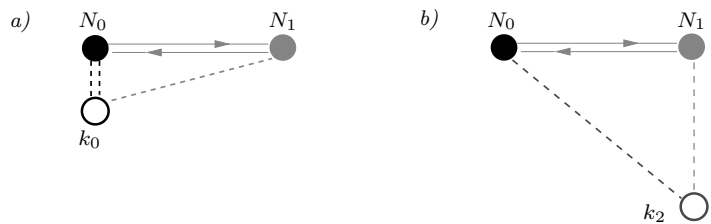
\begin{figure}[hbt]
\begin{center}
\begin{picture}(0,0)%
\includegraphics{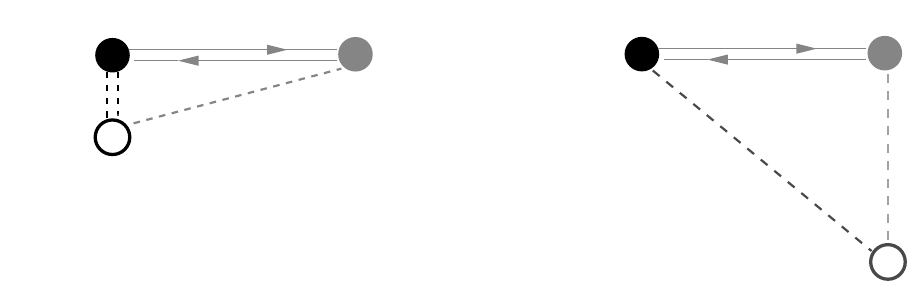}%
\end{picture}%
\setlength{\unitlength}{1381sp}%
\begingroup\makeatletter\ifx\SetFigFontNFSS\undefined%
\gdef\SetFigFontNFSS#1#2#3#4#5{%
  \reset@font\fontsize{#1}{#2pt}%
  \fontfamily{#3}\fontseries{#4}\fontshape{#5}%
  \selectfont}%
\fi\endgroup%
\begin{picture}(12447,3931)(-344,-3557)
\put(6946, 59){\makebox(0,0)[lb]{\smash{{\SetFigFontNFSS{8}{9.6}{\familydefault}{\mddefault}{\updefault}\emph{b)}}}}}
\put(901,-2086){\makebox(0,0)[lb]{\smash{{\SetFigFontNFSS{8}{9.6}{\familydefault}{\mddefault}{\updefault}$k_0$}}}}
\put(-329, 29){\makebox(0,0)[lb]{\smash{{\SetFigFontNFSS{8}{9.6}{\familydefault}{\mddefault}{\updefault}\emph{a)}}}}}
\put(901, 14){\makebox(0,0)[lb]{\smash{{\SetFigFontNFSS{8}{9.6}{\familydefault}{\mddefault}{\updefault}$N_0$}}}}
\put(4276, 14){\makebox(0,0)[lb]{\smash{{\SetFigFontNFSS{8}{9.6}{\familydefault}{\mddefault}{\updefault}$N_1$}}}}
\put(10936,-3421){\makebox(0,0)[lb]{\smash{{\SetFigFontNFSS{8}{9.6}{\familydefault}{\mddefault}{\updefault}$k_2$}}}}
\put(8161, 29){\makebox(0,0)[lb]{\smash{{\SetFigFontNFSS{8}{9.6}{\familydefault}{\mddefault}{\updefault}$N_0$}}}}
\put(11536, 29){\makebox(0,0)[lb]{\smash{{\SetFigFontNFSS{8}{9.6}{\familydefault}{\mddefault}{\updefault}$N_1$}}}}
\end{picture}%
\end{center}
\caption{D3/D$(-1)$-quiver for SQCD with \emph{a)} gauge instantons and \emph{b)} stringy instantons. 
Filled and empty circles represent stacks of D3 and D$(-1)$ branes,
solid lines stand for chiral bifundamental matter, dashed lines for charged instanton moduli.
}
\label{fk0k2}
\end{figure}

Some gauge and stringy instanton quiver diagrams are displayed in Fig. \ref{fk0k2}.

This leads to a very rich structure of non-perturbative interactions that include the
holomorphic ADS superpotential when $N_f=N_c-1$ and the multi-fermion F-terms of the
Beasley-Witten (BW) type \cite{Beasley:2004ys} when $N_f\geq N_c$, plus their possible
multi-instanton extensions.

We have seen that by computing the mixed open/closed string diagrams 
in the $\mathbb Z_2 \times \mathbb Z_2$ orbifold, one finds that
the flux induced interactions on the instanton moduli space
are encoded in the action 
\begin{equation}
S^{\mathrm{flux}} = {2\pi\ii}\, \left[
\frac{2G_{(3,0)}}{\sqrt{g_s}}\,\theta^{\alpha} \theta_{\alpha}
-\frac{2G_{(0,3)}}{\sqrt{g_s}}\,\frac{\pi^2{\alpha'}^2}{2}\,\lambda_{\dot\alpha}\lambda^{\dot\alpha}
\right]
+{\ii}\sqrt{g_s}\,G_{(3,0)}\,{\bar\mu}\mu~.
\label{sflux1}
\end{equation}
where we have inserted the correct normalization factors in (\ref{sflux01}).
Note that in the first and last
terms of (\ref{sflux1}) the scaling dimension of (length)$^{-1}$ carried by the $G$-flux is 
compensated by the dimensions of $\theta$, $\mu$ and $\bar\mu$, while in the second term explicit $\alpha'$ factors are needed. This is perfectly consistent with the fact that, while the $G_{(3,0)}$ flux components have a natural field theory interpretation as gaugino masses, the 
$G_{(0,3)}$ components instead have no counterpart on the gauge field theory. Thus, from the open
string point of view their presence in (\ref{sflux1}) is a genuine string effect, 
as revealed by the explicit factors of $\alpha'$.

The action (\ref{sflux1}) can be conveniently rewritten as
\begin{equation}
S^{\mathrm{flux}} = \frac{2\pi\ii}{g_s}\, \Big[\,G\,\theta^{\alpha} \theta_{\alpha}
-\bar G\,\frac{\pi^2{\alpha'}^2}{2}\,\lambda_{\dot\alpha}\lambda^{\dot\alpha}
\Big]
+\frac{\ii}{2}\,G\,{\bar\mu}\mu~,
\label{sflux2}
\end{equation}
where we have defined
\begin{equation}
 G = 2\sqrt{g_s} \,G_{(3,0)}~,\quad
\bar G = 2\sqrt{g_s} \,G_{(0,3)}~.
\label{GbarG}
\end{equation}
Eq. (\ref{sflux2}) is the form of the flux induced moduli action which we
will use in the following to study the non-perturbative interactions generated by
fractional D-instantons in the presence of bulk fluxes.
In particular we will consider terms at the linear order in $G$ or $\bar G$ where the world-sheet derivation of the moduli action (\ref{sflux2})
are reliable.
We therefore have two possibilities depending on whether we keep $G$ or $\bar G$
different from zero, which we are going to analyze in turn (see \cite{Billo':2008pg} for the details).

\paragraph{One-instanton effects with $G\neq 0$}
\label{subsecn:Gflux}
In this case we can set $\bar G=0$ and look for the non-perturbative interactions proportional
to $G$, assuming that the fractional D-instanton is of type 0, {\it i.e.} that $k_0=1$.
A class of such interactions is obtained by exploiting the $\frac{\ii}{2}G\bar\mu\mu$ term of the
flux action (\ref{sflux2}). At first order in $G$ this leads to
\begin{equation}
S_{\mathrm{n.p.}}(G) = 
\Lambda^{\beta_0}\int d^4x\,d^2\theta
\,d\,{\widehat{\mathfrak M}}
~\ee^{-S_{\mathrm{D3/D(-1)}}(\Phi,\bar\Phi)}\,\left(\frac{\ii}{2}\,G\bar\mu\,\mu\right)
\label{seffg}
\end{equation}
where $S_{\mathrm{D3/D(-1)}}(\Phi,\bar\Phi)$ is the instanton action. $\Lambda$ is the dinamically generated scale of the effective SU($N_0$) SQCD theory we are considering:
\beq
\Lambda^{\beta_0} = M_s^{\beta_0} e^{2\pi i \tau_0}~.
\eeq
By taking the field theory limit $\alpha'\to 0$ we
obtain non-perturbative flux-induced terms in the effective action of the form
\begin{eqnarray}
S_{\mathrm{n.p.}}(G) &=& \int d^4x\,d^2\theta \,{W}_{\mathrm{n.p.}}(G)
~,\\
{W}_{\mathrm{n.p.}}(G) &=& \Lambda^{\beta_0}
\int d\,{\widehat{\mathfrak M}}
\,~\ee^{-S^{(0)}_{\mathrm{D3/D(-1)}}(\Phi,\bar\Phi)}\,\left(\frac{\ii}{2}\,G\bar\mu\,\mu\right)
\label{weffg}
\end{eqnarray}
where $S^{(0)}_{\mathrm{D3/D(-1)}}(\Phi,\bar\Phi)$ is the moduli action in the field theory
limit.

The case  $N_1=N_0-1$ is particularly interesting, 
since it corresponds to $\mathrm{SU}(N_c)$ SQCD with $N_f=N_c-1$. 
In this case, in absence of flux,  the
gauge instanton induces the ADS superpotential; now we see that in presence of a $G$-flux
which softly breaks supersymmetry by giving a mass to the gaugino, the gauge instanton produces
new types of low-energy effective interactions.  By explicitly performing the integral
over the instanton moduli for $\mathrm{SU}(2)$ SQCD with
one flavor,  the flux induced non-perturbative term can be written as
\begin{equation}
{W}_{\mathrm{n.p.}}(G)= \left. \mathcal C\,G\,\Lambda^5\,\frac{\bar D^2 \bar M}{(\bar M M)^{3/2}}
\right|_{\bar\theta=0}
\label{N2G}
\end{equation}
where $M$ is the meson superfield of the effective theory and we made explicit the fact that no $\bar\theta$ dependence arises, due to the half-BPS nature of the D3/D(-1) system. We can regard this interaction
as a low-energy non-perturbative effect of the soft supersymmetry breaking realized by the
$G$-flux in the microscopic high-energy theory.
Finally, we observe that one can alternatively exploit the
$G\theta^2$ term of the  flux action to produce non-supersymmetric
interactions of the same type as the ones described here.

\paragraph{One-instanton effects with $\bar G \neq 0$}
\label{subsecn:barGflux}
The contribution to the effective action linear in $\bar G$ in presence of a single fractional D-instanton of type 0 is given, in analogy to Eq. (\ref{seffg}), by
\begin{equation}
S_{\mathrm{n.p.}}(\bar G) = \Lambda^{\beta_0}\,\int d^4x\,d^2\theta
\,d^2\lambda\,\,d\,{\widehat{\mathfrak M}'}
~\ee^{-S_{\mathrm{D3/D(-1)}}(\Phi,\bar\Phi)}\,
\left(-\frac{2\pi\ii}{g_s}\,\frac{(\pi\alpha')^2}{2}\,\bar G \lambda_{\dot\alpha}\lambda^{\dot\alpha}\right)~.
\label{seffgbar}
\end{equation}
Here we have denoted by ${\widehat{\mathfrak M}'}$ all centered moduli but $\lambda$. 
Performing the Grassmannian integration over $d^2\lambda$, we can write
\begin{equation}
S_{\mathrm{n.p.}}(\bar G) = \int d^4x\,d^2\theta
\,\,W_{\mathrm{n.p.}}(\bar G)~,
\label{seffgbar2}
\end{equation}
where
\begin{equation}
W_{\mathrm{n.p.}}(\bar G)=(\pi\alpha')^2 \,\frac{2\pi\ii}{g_s}\,\Lambda^{\beta_0}\,\bar G\, 
\int d\,{\widehat{\mathfrak M}'} \,~\ee^{-\left.S'_{\mathrm{D3/D(-1)}}(\Phi,\bar\Phi)\right.
}
\label{weffgbar}
\end{equation}
with $S'_{\mathrm{D3/D(-1)}}(\Phi,\bar\Phi)$ being the action without the fermionic
ADHM constraint term since the Grassmannian integration over $\lambda$ has killed it. 

Let us focus on the simple case $N_0=N_1$, which corresponds to a SQCD with $N_f=N_c$ flavors. 
In absence of fluxes, one gets multi-fermion terms of Beasley-Witten type. Here, for $N_c=2$, the result is
\begin{equation}
 \label{GbarN2}
{W}_{\mathrm{n.p.}} = \mathcal{C} \,\alpha'^2\, \bar G\,\Lambda^{4}\,
\left.\frac{\det \bar M}{\big(\tr \bar M M+\bar B B+ \bar {\tilde B}\tilde B\big)^{1/2}} \right|_{\bar\theta=0}~,
\end{equation}
where $M$ is the meson superfield and $B$ and $\widetilde B$ are the baryon superfields.

\paragraph{Stringy instanton effects in presence of fluxes}
\label{secn:strinst}
D-instantons of type $2$ and $3$ are of different type with respect to
the D3 branes where the SQCD-like $\mathrm{SU}(N_0)\times \mathrm{SU}(N_1)$ theory
is defined, and lead thus to exotic non-perturbative effects.
The moduli action  drastically
simplifies. In particular, it  does not contain any $\lambda$ dependence.
Therefore, unless one introduces an orientifold projection
or invokes other mechanisms,
the only way to get a non-zero result is to include
the flux-induced $\bar G\lambda\lambda$ term of Eq. (\ref{sflux2}) and
use it to perform the $\lambda$ integration. At the linear level in the fluxes,
the other flux interactions in (\ref{sflux2}) become then irrelevant.
Neglecting as usual numerical prefactors, we can write the centered partition function for
a stringy instanton configuration with instanton numbers $k_2$ and $k_3$ as
\begin{equation}
 \label{Wsi}
W_{\mathrm{n.p.}}(\bar G) = \alpha'^2\,\bar G\,M_s^{k_2\beta_2 +
k_3\beta_3}\,\ee^{2\pi\ii(k_2\tau_2 + k_3\tau_3)}\int d\,{\widehat{\mathfrak
M}'}\, \ee^{-S_{\mathrm{D3/D(-1)}}}~.
\end{equation}
Let us now concentrate on the set-up containing a single stringy instanton
described in Fig. \ref{fk0k2}\emph{b)}, namely let us set $k_2=1$, $k_3=0$.
As remarked above, 
the only way to saturate the Grassmannian integration over
$d\lambda_{\dot\alpha}$ is via their $\bar G$ interaction
and the non-perturbative contribution to the effective action of this 
``stringy'' instanton sector is
\begin{equation}
 \label{Seffsi}
S_{\mathrm{n.p.}} = 
\int d^4x\,d^2\theta\, W_{\mathrm{n.p.}}(\bar G)
\end{equation}
where the superpotential is given by
\begin{equation}
{W_{\mathrm{n.p.}}}= \mathcal C\,{\alpha'}^2\,
M_s^{-(N_0 + N_1)}\,\ee^{2\pi\ii\tau_2}\,\bar G\,
\int d\,{\widehat{\mathfrak M}'}
\,\ee^{-S_{\mathrm{D3/D(-1)}}(\Phi)}~.
\label{Weffsi}
\end{equation}
Notice that the dimensional prefactor does not combine
with the exponential of the classical action  to form the
dynamically generated scale of the gauge theory, since
$\tau_2$ is the complexified coupling of D3-branes of type 2, which are not the ones that
support the gauge theory we are considering.

Performing the integral over the moduli, we find that a single
stringy instanton in presence of an imaginary self-dual three-form flux produces for  $N_f=N_c$ flavors a holomorphic superpotential
\begin{equation}
 \label{risweffsi}
W_{\mathrm{n.p.}}= \mathcal{C}\, M_s^{2 - 2
N_c}\ee^{2\pi\ii\tau_2}\, \bar G\, \det M~.
\end{equation}
Interestingly, the interactions generated by stringy instantons
are still holomorphic and therefore supersymmetric even in the
presence of the supersymmetry breaking flux $\bar{G}$.

\section{Conclusions}

With this note we hope to have convinced the reader that closed string background fluxes and instantons in string theory are actual and very interesting topics, which  lead to useful phenomenological applications. Indeed, on one side, their interplay modifies the field theory results on superpotentials, giving rise to new types of low-energy effective actions. Moreover, they provide a mechanism to get non-vanishing contributions from exotic instantons, which represent an useful tool to be used in the search for a stringy description of phenomenological models.

\acknowledgments
I would like to thank  Marco Bill\`{o}, Marialuisa Frau,  Francesco Fucito, Alberto Lerda and Jose Francisco Morales for many illuminating discussions. I thank INFN and SIF for having given me the opportunity of writing this note.
L. F. is supported by CNRS and LAPTH in Annecy-le-Vieux.


\end{document}